\documentclass{elsart}
\usepackage{psfig}
\begin{document}

\begin{frontmatter}
\title{Phase Transition of DNA-Linked Gold Nanoparticles}

\author{Ching-Hwa Kiang}
\address{Department of Physics \& Astronomy,
Rice University, Houston, TX\ \ 77005}

\begin{abstract} 
Melting and hybridization of DNA-capped gold nanoparticle networks are
investigated with optical absorption spectroscopy.
Single-stranded, 12-base DNA-capped gold
nanoparticles are linked with complementary, single-stranded, 24-base
linker DNA to form particle networks.  Compared to free DNA, a sharp melting 
transition is seen in these networked DNA-nanoparticle systems.
The sharpness is explained by percolation transition phenomena.
\end{abstract}
\end{frontmatter}

%\pacs{87.15.-v, 87.64.-t, 87.64.Ee}

%87.15.-v Biomolecules: structure and physical properties
%87.64.-t Spectroscopic and microscopic techniques in biophysics and medical physics
%87.64.Ee Electron microscopy

%%%%%%%%%%%%%%%%%%%%%%%%%%%%%%%%%%%%%%%%%%%%%%%%%%%%%%%%%%%%%%%%%%%%%%%%

\section{Introduction}

DNA melting and hybridization phenomena are of great importance in
both fundamental biological science and biotechnology
\cite{wartell85a,hwa97a,breslauer99a,libchaber00a}.  Sequence-specific
DNA recognition is important in detection, diagnosis of genetic
diseases, and identification of infectious agents.  
DNA-gold nanoparticle
systems undergo a change of color upon network formation, which can be
used for highly sensitive detection when specific nucleic acid
sequences induce network formation \cite{Mirkin98a}.
Indeed, such a system has been demonstrated to be useful to detect
anthrax and other agents of biowarfare \cite{Mirkin00d}.

Here we report studies of DNA-gold nanoparticle melting transitions.
DNA-gold nanoparticles were prepared using modified methods
based on Ref. \cite{Mirkin98a}.  The basic unit is illustrated in
Fig.~\ref{fig1}.  Gold nanoparticles are derivatized with
non-complementary, single-stranded, 12-base DNA ($\sim$4~nm).  Upon adding 
24-base, complementary linker DNA, particles aggregate to form network
structures, and precipitate out of the solution.

DNA-functionalized gold nanoparticles were prepared by conjugating
gold nanoparticles ranging from 10 to 40~nm in diameter (Sigma, ICN
Pharmaceutical) with thiol-modified DNA \cite{Mirkin98a,Chrisey96a}.
3'-alkanethiol DNA 3'S-(CH$_2$)$_3$-ATG-CTC-AAC-TCT was prepared by
using \mbox{C3 S-S} modifier and 5'-alkanethiol DNA
TAG-GAC-TTA-CGC-(CH$_2$)$_6$-S5' by \mbox{C6 S-S} on a 1 $\mu$mol
scale and purified by HPLC (Invitrogen).  Just prior to conjugation
with gold, 100 nmol of the dried DNA was redispersed in 400 $\mu$l of
0.1 M DTT, 0.1 M phosphate buffer (pH~8) solution at room temperature
for 30 minutes to cleave the disulfide bond.  Salt was removed with a
NAP-5 desalting column to avoid bare gold aggregation prior to DNA
conjugation.

The deprotected alkanethiol-modified DNA were used to derivatize gold
nanoparticles at room temperature for 24 hours to form gold
nanoparticle probes.  The solutions were then brought to 0.3~M NaCl,
10~mM phosphate buffer (pH~7) and allowed to stand for 48 hours.  To
remove excess DNA, solutions were centrifuged at 13,200~$rpm$
(16,110$\times G$) for 60~minutes.  The supernatant was decanted, and
the red precipitate was redispersed in 1~ml nanopure water and
centrifuged again.  After decanting the supernatant, about 200~$\mu$l
of precipitate of each modified gold nanoparticles were collected
for spectroscopic investigation.
 
20~nmol linker DNA 3'GCG-TAA-GTC-CTA-AGA-CTT-GAG-CAT5' (Invitrogen)
was dispersed in 1~ml solution of 0.3 M NaCl, 10 mM phosphate buffer
(pH~7).  Hybridization of linkers with gold nanoparticles was done by
mixing 200~$\mu$l each modified gold nanoparticles (0.8~OD) with
8~$\mu$l linker solution (10~pmol/$\mu$l).  The solution was
annealed at 70~$^\circ$C for 10 minutes and cooled to room temperature
during a 2 hour period and was allowed to aggregate for
several days.

Absorption spectra of DNA-modified gold nanoparticles were taken
on a Hewlett-Packard diode array spectrophotometer (HP8453).
The kinetics of formation of network nanoparticle structures at 
room temperature is illustrated in Fig.~\ref{fig2}.
Upon adding linker DNA, gold nanoparticles aggregate to form networks,
as demonstrated in the gold surface plasmon peak (520~nm) shift of the
DNA-modified gold nanoparticles \cite{Schatz00a,Elsayed99a}.  
The aggregation started with the wavelength shift of
the plasmon band, followed by broadening and more shifting of the peak
as hybridization continues.  From molecular simulations it is
known that change in network size results mainly in peak broadening, 
whereas change in gold volume fraction results mainly in peak 
shifting \cite{Mirkin00a}.  Our results indicate that the initial 
aggregation has characteristics consistent with increasing volume 
fraction, followed by increasing network size.

To study the equilibrium behavior of DNA-gold nanoparticle system, we
monitor the UV-visible spectra absorption intensity at 260~nm and
520~nm while melting the DNA-gold nanoparticle network.  DNA bases
have strong absorption at the UV region, and the peak near 260~nm is a
result of combination of these electronic transition dipoles
\cite{CantorII}.  The DNA double helix has smaller extinction 
coefficient than single-stranded DNA due
to hypocromism and, therefore, the absorption intensity at 260~nm
increases as a result of DNA melting.  The sample was heated by a
peltier temperature controller (JASCO J-715) at a rate of 0.5
$^\circ$C/min, from 25 to 75 $^\circ$C.  The 260~nm and 520~nm melting
curves are very similar, indicating that DNA and nanoparticle melting
are closely related.  Figure~\ref{fig3}a displays the melting
curves of 10~nm, 20~nm, and 40~nm gold particles with linker DNA.
The melting transition width (FWHM) is about 5$^\circ$C, compared to
12$^\circ$C for melting of free DNA \cite{Mirkin97a}.  As illustrated
in the figure, the transition width as well as the melting temperature
$T_m$ of DNA have been dramatically modified by the binding to gold
particles.  

Figure~\ref{fig3}b illustrates two different transitions at high
$T$ and low $T$.  The absorption of DNA-modified, 10~nm, 20~nm, and
40~nm gold nanoparticles are plotted versus reduced temperatures
($T_R=T/T_m$).  Above temperature $T_m$ (low connectivity),
the curves appear insensitive to details,
indicative of universal scaling at the percolation
transition \cite{Rudnick98a}.  The curves can be fitted with an
equation that describes percolation phenomena,
$$
A(T_R) = 1-a (T_c-T_R)^\beta
%y=1-4.07877*(1.00486-x)^0.403
$$
where $A(T_R)$ is the absorption as a function of reduced temperature
$T_R$, and $\beta$ is the critical exponent for percolation.  In
three-dimensions, $\beta=0.403$ \cite{Creswick92a}.  The parameters
$a$ and $T_c$ are adjustable.  Using the data for all three particle sizes
for $1<T_R<1.003$, we obtained $a$=4.079 and $T_c$=1.005, as shown in
Fig.~\ref{fig3}b.  The deviation from the fit above $T_c$ is
presumably due to finite size effects \cite{Deem95a}.

The growth mechanism of the DNA-gold nanoparticle network is illustrated in
Figure~\ref{fig4}.  Particles initially are dissolved in the
solution.  With the addition of complementary linker DNA, hybridization
occurs, and the particles form a network-like structure.  The volume
fraction of the porous structure continues to increase past the 
percolation threshold, and eventually the clusters become a 
dense amorphous structure.

In conclusion, the network formation in DNA-gold particle systems
exhibits a phase transition, which does not occur for short DNA.  Our
results provide detailed and precise measurements of
hybridization/melting of the DNA-gold nanoparticle system.  DNA
attached to gold is a controlled system in which to do experiments 
on phase transitions and serves as a good probe for many biological systems
that involve DNA.  This is a new system, about which relatively little
is known.  Many parameters, such as the length of DNA, particle size,
degree of disorder, solution pH value, and salt concentration, 
may be varied to better understand the system.

$^*$To whom correspondence should be addressed, 
Email: chkiang@rice.edu.

\bibliography{au1}

\clearpage
\newpage

\begin{figure}[p]
\caption[]
{\label{fig1}
Basic building block of gold nanoparticles capped with
12-base DNA with thiol-modification at 3'($D1$) or 5'($D2$) end.  $L$
is
linker DNA composed of 24-base DNA complementary to the
capping DNA $D1$ and $D2$.  Capping DNA is bound to gold nanoparticles
through covalent bonds, and the linker DNA $L$ binds the capping
DNA through hydrogen bonds.
}
\end{figure}

\begin{figure}[p]
\caption[]
{\label{fig2}
Optical absorption spectra of DNA-modified gold nanoparticles.
The time starts when linker $L$ is added to the solution containing
mixtures of gold nanoparticles modified with $D1$ and $D2$.
Owing to the particle network formation, the 520~nm gold surface
plasmon peak slowly shifts to longer wavelength, followed by peak
broadening and further shifting.}
\end{figure}

\begin{figure}[p]
\caption[]
{\label{fig3}
Normalized melting curves of linked gold nanoparticle networks
monitored at 260~nm.  (a) Melting curves of different size particles
as a function of temperature.  Inset shows the change in melting
temperature, $T_m$, with particle diameter $D$.  The line is
a fit to a power law. (b) Melting of particles as a function of reduced
temperature.  Solid line shows the region where data were used for
curve fitting, dashed line is calculated from the equation of
best fit (see text).  The scaling shows the universality of the
melting transition above the transition temperatures.
}
\end{figure}

\begin{figure}[p]
\caption[]
{\label{fig4}
Growth mechanism of the DNA-gold nanoparticle network.  When linker
DNA is added, hybridization occurs, connecting the gold nanoparticle
into porous networks.  As the cluster grows, the volume fraction and
the size of the cluster continue to increase.  The transition between
dispersed nanoparticles and micrometer sized cluster appears to be a
cooperative phenomenon \cite{Bruinsma99a}, much like a phase
transition.
}
\end{figure}

\clearpage

\newpage
\psfig{file=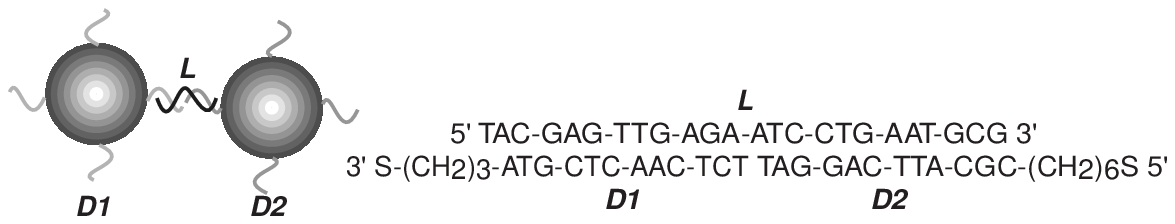,width=5in,angle=0}
\vfill
Figure \ref{fig1}. Kiang, ``Phase Transition \ldots.''

\newpage
\psfig{file=fig2.eps,width=5in,angle=0}
\vfill
Figure \ref{fig2}. Kiang, ``Phase Transition \ldots.''

\newpage
\psfig{file=fig3.eps,width=5in,angle=0}
\vfill
Figure \ref{fig3}. Kiang, ``Phase Transition \ldots.''

\newpage
\psfig{file=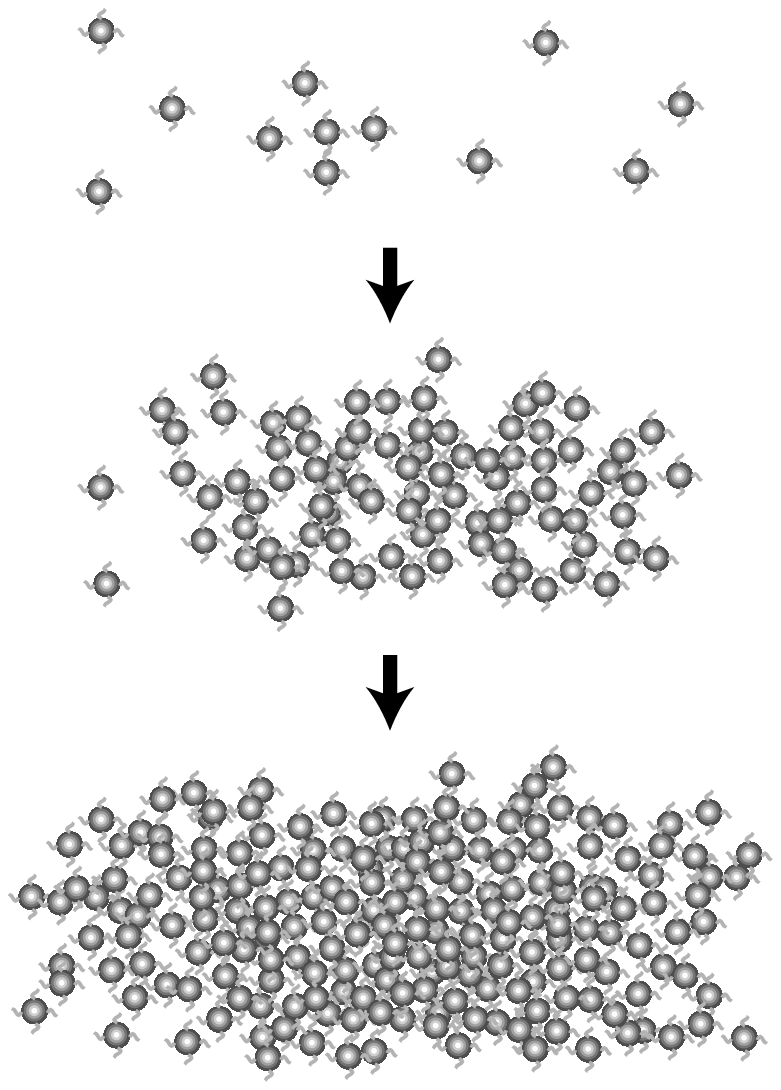,width=5in,angle=0}
\vfill
Figure \ref{fig4}. Kiang, ``Phase Transition \ldots.''

\end{document}